# International Migration in Academia and Citation Performance: An Analysis of German-Affiliated Researchers by Gender and Discipline Using Scopus Publications 1996-2020

Xinyi Zhao[1], Samin Aref[1], Emilio Zagheni[1], and Guy Stecklov[2]

[1] *{zhao, aref, zagheni}@demogr.mpg.de*
Laboratory of Digital and Computational Demography, Max Planck Institute for Demographic Research, Konrad-Zuse-Str. 1, 18057 Rostock (Germany)

[2] *guy.stecklov@ubc.ca*
Department of Sociology, University of British Columbia, 6303 NW Marine Drive, Vancouver, BC (Canada)

**Abstract**
Germany has become a major country of immigration, as well as a research powerhouse in Europe. As Germany spends a higher fraction of its GDP on research and development than most countries with advanced economies, there is an expectation that Germany should be able to attract and retain international scholars who have high citation performance. Using an exhaustive set of over eight million Scopus publications, we analyze the trends in international migration to and from Germany among published researchers over the past 24 years. We assess changes in institutional affiliations for over one million researchers who have published with a German affiliation address at some point during the 1996-2020 period. We show that while Germany has been highly integrated into the global movement of researchers, with particularly strong ties to the US, the UK, and Switzerland, the country has been sending more published researchers abroad than it has attracted. While the balance has been largely negative over time, analyses disaggregated by gender, citation performance, and field of research show that compositional differences in migrant flows may help to alleviate persistent gender inequalities in selected fields.

**Introduction**

In the current era of knowledge-based economies, highly skilled people are the most mobile population group worldwide (Schiller & Cordes, 2016). In OECD countries, highly qualified individuals, including researchers, make up one-third of the immigrant population (Docquier & Marfouk, 2004; Schiller & Cordes, 2016). The international mobility of researchers facilitates the exchange of knowledge, ideas, and skills, and thus contributes to the dynamic development of the global knowledge production system (Bauder, 2015; Franzoni et al., 2015; Netz & Jaksztat, 2017). Countries see researchers as prized groups because of their high levels of human capital and potential for fueling innovation and economic growth. Gaining a better understanding of this kind of mobility is crucial for evaluating scientific research at the national level, and for informing policy on science and mobility in academia (Netz et al., 2020). As Germany is a science powerhouse that hosts more non-EU researchers than any other EU country (Jöns, 2009; IDEA Consult, 2013; Lörz et al., 2016; Guthrie et al., 2017; Aman, 2018), Germany is expected to be able to attract and retain international researchers with high citation performance. A lack of fine-grained data on the international migration of German-affiliated researchers and their citation performance makes it difficult to understand the in- and out-flow patterns and costs associated with this kind of mobility. This study develops individual-level migration data in order to analyze the mobility of researchers to and from Germany, while taking into account the citation performance of internationally mobile scholars across disciplines and genders.

Germany's relatively liberal immigration laws and its powerful position as the world's fourth-largest economy (Schiller & Cordes, 2016) are key factors in explaining why about one-fifth of the top-level researchers in Germany in 2004 were foreign-born (Ioannidis, 2004). Over the past three decades, German universities and research institutions have attracted large numbers of researchers from other countries, especially from other European countries (Giousmpasoglou

& Koniordos, 2017). In addition, Germany spends over 3.1% of its gross domestic product (GDP) on research and development (R&D), which is more than the R&D spending by GDP in most OECD countries, including the United States (US) (2.8%) and the United Kingdom (UK) (1.7%). It has been suggested that highly cited researchers tend to migrate to countries with higher R&D spending (Hunter et al., 2009). However, when the two migration directions are considered, Germany appears to have suffered a net loss of 28% due to the global migration of researchers (Schiller & Cordes, 2016; OECD, 2008). More specifically, Germany is experiencing a "brain drain" in certain specializations, including medical research (Giousmpasoglou & Koniordos, 2017). Although existing studies on the German science system have explored some aspects of academic migration (Netz & Finger, 2016; Aman, 2016; Parey et al., 2017; Netz & Jaksztat, 2017; Netz & Grüttner, 2020), there has been no systematic analysis of international academic migration for Germany that has taken the citation performance of researchers into account, and that has covered all fields of scholarship. This study provides a much-needed baseline for the development of future policies that can help Germany succeed in attracting and retaining qualified researchers from overseas, facilitating the circulation of brainpower, and advancing the performance of the German science system.

It is generally recognized that international academic mobility has a positive influence on the global scientific system. By promoting knowledge production and diffusion between countries, migration among researchers enhances the performance of global science production (Giousmpasoglou & Koniordos, 2017; Guthrie et al., 2017; Netz et al., 2020). From a micro perspective, there is strong evidence that mobile researchers outperform non-mobile researchers (Dubois et al., 2014; Franzoni et al., 2014; Moed & Halevi, 2014; Guthrie et al., 2017). Findings from the MORE survey have shown that mobility leads to increased outputs for researchers in both academia and industry. Notably, there is evidence that the output effects are higher among researchers moving from the EU to the US than among those moving in the opposite direction (Børing et al., 2015). Gibson and McKenzie found that emigrant researchers have much greater research outputs and impacts, as measured by total citation and $h$-index, than researchers who stay in their respective origin countries (Gibson & McKenzie, 2014). A comprehensive evaluation of the interplay between migration and citation performance that takes individual all disciplines into account can deepen our understanding of this crucial topic.

Qualitative interviews, surveys, bibliometric data, and data from curricula vitae are among the most common data sources for studying the migration of researchers (Netz & Jaksztat, 2017; Netz & Grüttner, 2020). The idea of using the historical records of researchers to follow their geographical movements can be traced back to a study by Rosenfeld & Jones (1987) on the movements of psychologists in the US that used a sample from the biographies of members of the American Psychology Association. The digital revolution and the advent of digitized sources of bibliometric information enables us to expand this simple idea to cover a large number of data points with a flexible level of granularity that is suitable to our research objectives. Previous studies that utilized similar applications of bibliometric data have mapped academic mobility among countries, and have shed light on the causes and consequences of academic mobility (Moed & Halevi, 2014). Recent methodological innovations for re-purposing bibliometric data to study internal migration within country boundaries have made the process of inferring migration events from affiliation addresses more reliable (Miranda-González et al., 2020).

This paper relies on large-scale digitized bibliometric data from *Scopus* to document and analyze international migration to and from Germany among researchers over the past 24 years. The cleaned and pre-processed bibliometric data provide a unique perspective on the international migration patterns and geographical trajectories of mobile researchers. The results help to clarify the position of Germany in the global science system. In addition, the analysis evaluates the interplay between migration and citation performance across different disciplines.

Thus, this analysis adds a demographic dimension to the science of science literature by providing several in-depth statistics related to academic mobility.

**Materials and methods**

*Scopus publications of all German-affiliated authors*

This paper relies on a complete enumeration of the Scopus-indexed publications of researchers who have published with an affiliation address in Germany at some point during the 1996-2020 period (over eight million publications from more than one million researchers). Note that these data do not provide information about researchers who have not published in Scopus-indexed sources during the temporal window. The unit of the data is an *authorship record,* which we define as the linkage between an author affiliation and a publication.

The raw data are pre-processed to ensure that we are performing a reliable analysis of mobility events based on the changes in affiliation addresses. The pre-processing steps involve the adoption of an unsupervised machine learning algorithm for disambiguating authors and a neural network algorithm for handling missing values (Miranda- González et al., 2020; Subbotin & Aref, 2020). In our dataset, a large majority of authorship records have a country variable. However, there are 96,465 authorship records with missing country information. The missing values are systematically inferred using a neural network algorithm inspired by Miranda-González et al. (2020), which takes an affiliation address as the input and predicts the country as the output. We use a random set of one million authorship records from our dataset that have country information, and use them as training data (80%) and testing data (20%). The technical details of the development of the neural network have been explained elsewhere (Miranda-González et al. 2020). It has been demonstrated that the neural network can correctly predict the country for 98.4% of records, which is a level of accuracy we consider acceptable for predicting missing country information.

The second step of our data pre-processing helps us overcome the problems associated with using Scopus author IDs to identify unique authors. It has been shown that Scopus author IDs have high levels of *precision* and *completeness*. Precision measures the percentage of author IDs that are associated with the publications of a single individual only. Completeness measures the percentage of author IDs that are associated with all of the Scopus publications of an individual. The results of the latest evaluation of the accuracy of Scopus author IDs conducted in August 2020 showed that the precision and the completeness of Scopus author profiles are 98.3% and 90.6%, respectively (Paturi & Loktev, 2020). However, while it appears that the quality of individual-level Scopus data is sufficiently high to enable us to study the migration of researchers (Kawashima & Tomizawa, 2015; Aman, 2018), there are several notable limitations to keep in mind when using Scopus data for migration research. The precision limits in Scopus author IDs imply that 1.7% of Scopus author IDs may be associated with the publications of more than one person, which could affect the accuracy of the migration events detected through changes of affiliation countries per author ID. We overcome this problem by using an unsupervised machine learning algorithm that was inspired by a recently developed author name disambiguation method (D'Angelo and van Eck, 2020). The conservative adaptation of this method (Miranda-González et al., 2020) assumes that every two authorship records are from distinct individuals unless sufficient evidence is found to the contrary using a rule-based scoring approach and a clustering method. We first calculate the similarity score of each pair of two authorship records belonging to the same author ID. The similarity is measured based on author names, co-author names, subjects, funding information, and grant numbers. The author disambiguation algorithm makes all pairwise comparisons between authorship records with the same author ID, and creates a distance matrix based on similarities and dissimilarities in the aforementioned features for each pair of records. A clustering algorithm is then used to process the distance matrices, and to cluster similar authorship records. We then issue revised author IDs based on the resulting clusters. We use the *agglomerative clustering* algorithm from the scikit-learn Python library (Pedregosa et al., 2011) to cluster authorship records. This algorithm belongs to the family of hierarchical clustering methods. Supporting our conservative approach, it first places each record in its own cluster, and then merges pairs of clusters successively if doing so minimally increases a given linkage distance (Pedregosa et al., 2011). As well as being compatible with our conservative approach, agglomerative clustering has the advantage of offering us the flexibility to process any pairwise distance matrix.

Our author disambiguation procedures are based on the extraction of a subset of *suspicious* author IDs that are especially likely to be affected by the precision flaws of Scopus author IDs. An author ID is considered suspicious if it is associated with more than six countries or more than 292 publications (an average of more than one publication per month across a period of 24 years and four months). Based on these criteria, only 25,000 out of a total of 1.4 million author IDs are classified as suspicious. These author IDs are associated with 2,242,797 publications. After disambiguation, revised author IDs are issued for these records according to their clusters, and then they are merged with the rest of the data for the purpose of inferring migration events. The international mobility of researchers is determined by identifying the changes in the affiliation addresses of authors across different publications over time. To detect migration events more reliably, the most frequent (mode) country(ies) of affiliation is extracted for each researcher in each year. A migration event is considered to have happened only if the mode country of affiliation changes for the researcher across different years such that the previous mode country disappears. By aggregating the movements for each pair of origin and destination countries, we can estimate the international scholarly migration flows. Accordingly, the country of academic origin is defined as the mode country during the first year of publishing. Similarly, the country of destination is defined as the mode country of the most recent year of publishing.

Based on their migration events or lack thereof, and on their academic origins and destinations, researchers are assigned to one of the following six categories (mobility types) from the perspective of the German science system:
(1) Single-paper author (having only one publication);
(2) Non-mover (having multiple publications and Germany as the only mode country);
(3) Immigrant (origin: not Germany; destination: Germany);
(4) Emigrant (origin: Germany; destination: not Germany);
(5) Return migrant (origin: Germany; destination: Germany, with international migration); and
(6) Transient (origin: not Germany; destination: not Germany, but with Germany being among the researcher's mode countries at some point).

The annual net migration rate (NMR) is calculated based on the difference between incoming and outgoing flows in a year divided by the population of research-active scholars in that year, and then expressed as a per-thousand rate. The size of the population of research-active scholars in a given year is based on the number of researchers who have listed Germany as their country of affiliation on publications within a two-year vicinity of that year. A positive NMR value can be interpreted as indicating that more researchers are entering than leaving the country under analysis.

We define the *academic age* of researchers as the number of years since their first publication (as of 2020). We calculate the average annual citation rate for each researcher by dividing their total number of citations (as of April 2020) by their academic age. This measure allows us to compare the citation performance of researchers who have different levels of experience, but are in the same field. When comparing the citations of migrant researchers from different fields, we apply a second normalization by dividing the annual citation rate by the average annual citation rate of migrant researchers in that field.

We use the *All Science Journal Classification* (ASJC) codes to incorporate the fields and disciplines (subfields) of scholarship in our analysis. The ASJC codes are based on four fields (*Heath Sciences*, *Life Sciences*, *Physical Sciences*, and *Social Sciences*). These codes are further divided into 26 academic disciplines that provide a more granular level of information about the scholarship. The field and the discipline of each researcher are determined based on the frequencies of the ASJC codes in the individual's publications. We first compute the frequencies of each field and discipline in the authorship records of an individual, and then compare them using a Z-test (Subbotin & Aref, 2020) with the mean and the standard deviation of frequencies for that field or discipline among all researchers in our dataset. The main field and discipline of each researcher are determined based on the largest Z-score that exceeds a threshold of one. If neither of the Z-scores exceeds one, we categorize the researcher as *Multidisciplinary*.

The first names of researchers can be used to infer their gender (Larivière et al., 2013; Goldstein & Stecklov, 2016b, a; Dworkin et al., 2020). This study utilizes the *Genderize* library, a simple application programming interface (API) that provides the gender most commonly associated with a given first name from a dataset of over 114 million names. After performing basic text operations (like removing middle initials from the first name field), we obtained the gender for 1,046,873 author profiles in our dataset. For the remaining profiles, we manually searched for public author information to determine the gender by checking the individual's personal homepages, curricula vitae, online profiles, biographies in publications, and other online sources. Accordingly, the genders for 6,767 additional author profiles were determined manually. Finally, the gender of 1,053,640 author profiles (75.96%) were reliably determined by either algorithmic or manual gender detection. For our analyses that involve gender (e.g., measuring gender ratios), we set aside the 24.04% of author profiles whose gender cannot be determined either algorithmically or manually. The binary genders inferred and used in our analysis do not refer directly to the sex of the researchers, assigned at birth or self-chosen; nor do they refer to the socially assigned or self-chosen genders of the authors.

**Results**

We look at citation-based performance by discipline across different mobility types. We also provide descriptive results on common countries of origin and destination, and calculate the annual net migration rates of researchers for Germany. Furthermore, the discipline-normalized citation performance of the researchers is evaluated and compared for the common origin and destination countries. Finally, we provide detailed results on the gender ratios of researchers by discipline, and compare the gender ratios among mobile researchers and all researchers.

*Mobility types and annual citation rates by discipline*

Among the researchers who have Germany as a mode country for at least one year, 44.01% are single-paper authors, 42.08% are non-movers, and only 13.91% are internationally mobile. More details are provided in Table 1.

**Table 1. Number and proportion of researchers by mobility type.**

|  | Single-paper author | Non-mover | Immigrant | Emigrant | Return migrant | Transient |
|---|---|---|---|---|---|---|
| Count | 610,449 | 583,679 | 51,135 | 63,003 | 37,078 | 41,722 |
| % | 44.01% | 42.08% | 3.69% | 4.54% | 2.67% | 3.01% |

Figure 1 shows the average, median, and standard deviations of annual citation rates by discipline for researchers of each mobility type. There are substantial disparities in the annual citation rates of these researchers, as the large standard deviations indicate. As expected, we find that there are notable differences in the annual citation rates of researchers across different disciplines (Marmolejo-Leyva et al., 2015; Bedenlier, 2017; Horta et al., 2019). On average, researchers in the social sciences (colored in blue) receive fewer citations and researchers in the life sciences (green) have more citations than their counterparts in other fields.

When we look at differences in citation rates by mobility type, we can see that mobile researchers markedly outperform non-movers, even after separating single-paper authors. Compared to non-movers, mobile researchers have much higher annual citation rates for almost all disciplines. Thus, even though they represent a minority of the researchers in Germany, mobile researchers make substantial scientific contributions in terms of receiving citations. Among the internationally mobile researchers, transients and return migrants have higher annual citation rates than immigrants and emigrants. This pattern may be explained in part by the more complex mobility trajectories of transients and return migrants.

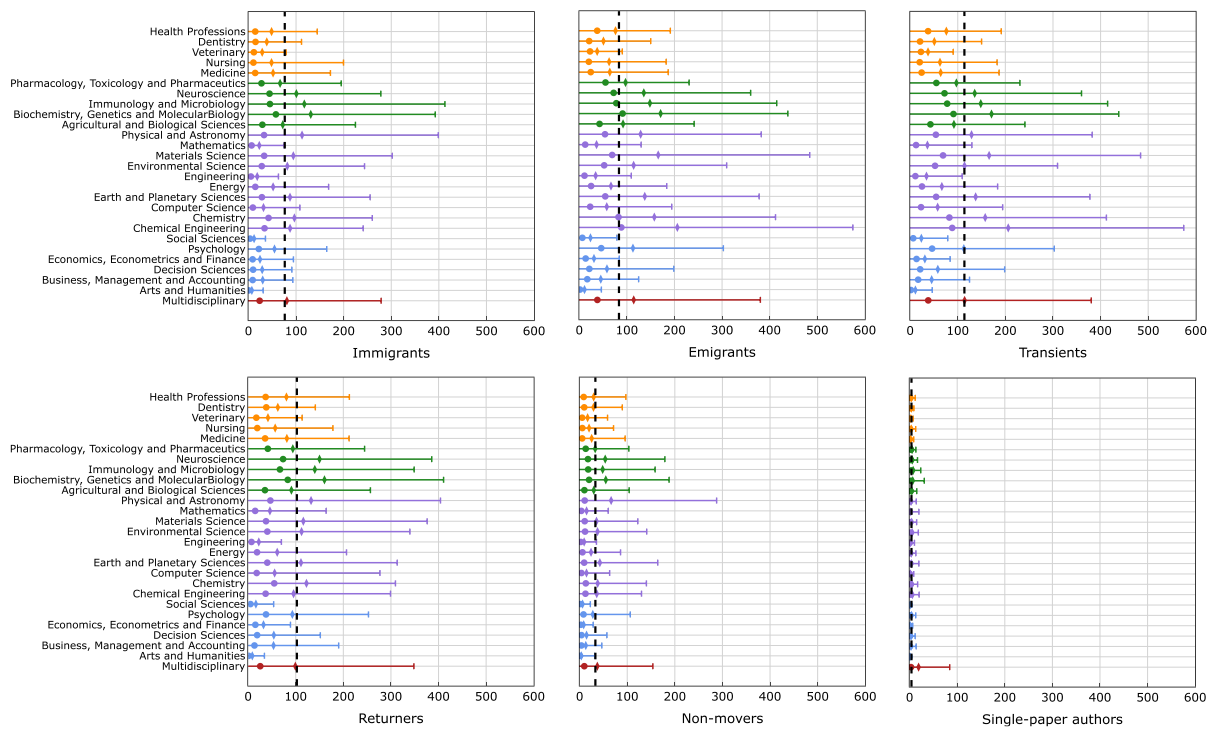

**Figure 1.** Average (diamond marker), median (circle marker), and standard deviation of annual citations by discipline and mobility type. The vertical dashed lines show the mean across all disciplines. Magnify all figures on the screen for higher resolution and more details.

*Flows, origins, and destinations*

Figure 2 illustrates the total migration flows of researchers from and to Germany during the 1996-2020 period. By far, the largest academic migration inflows and outflows are with the US, as various studies have documented (Maier et al., 2007; Weinberg, 2009; Hunter et al., 2009; Franzoni et al., 2012; Guthrie et al., 2017). For Germany, the US is also the most common resource hub of highly skilled individuals, due to the large number of researchers in the US and their high mobility potential (Schiller & Cordes, 2016). After the US, the common academic origins and destinations for Germany-affiliated migrant researchers are the UK in a distant second, followed by Switzerland. Most of the other common origin and destination countries are in Europe, which indicates that geographical proximity is also relevant. Moreover, EU policies might have played a role in facilitating large migration flows of scholars within Europe. For example, the *Bologna Process,* which was introduced in the 2000s, was designed to ensure comparability in the standards and the quality of higher education qualifications in Europe (Teichler, 2015). A more recent example is the *IPID4all* program, which was launched in 2014 with the aim of creating an attractive environment in Germany for young international researchers (IDEA Consult et al., 2017). Such polices are helping to drive and facilitate academic mobility by providing researchers with more opportunities for academic exchange and communication. In addition, there are a number of countries that are exporting substantial numbers of published researchers to Germany, but that are not equally attractive destinations for researchers from Germany, including Russia and India.

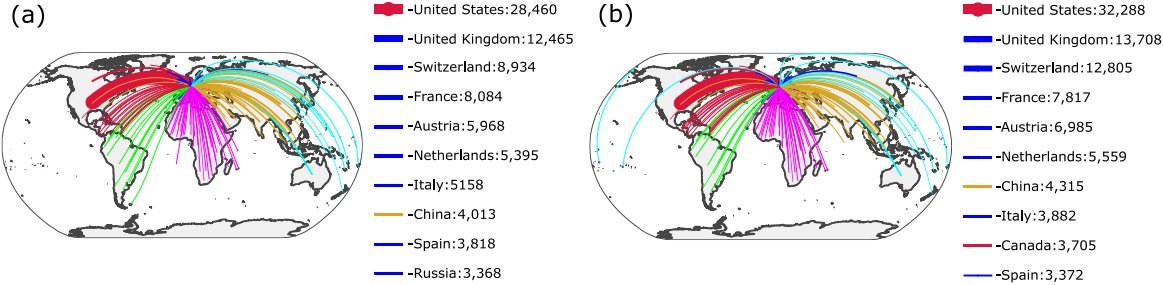

**Figure 2. Migration flows of researchers to Germany (a) and from Germany (b) over the 1996-2020 period.**

*Net migration rates*

Figure 3 illustrates the shifting annual net migration rates of researchers in Germany over the 1998-2017 period (other years are not reported due to boundary effects). The data suggest that for nearly the entire two decades of our analysis, the outgoing flow of published researchers exceeded the incoming flow. The lowest NMR value is for 2008, at -10.52 per 1,000 researchers. Note that the NMR in the general population also reached its lowest value in 2008, which was during the global economic crisis. However, in contrast to migration among researchers, the net migration rate for the general population as a whole was consistently positive throughout these years. After 2008, the NMR for researchers displayed a generally increasing trend until 2014, when it reached a peak of +1.56. In the most recent years, the NMR for researchers followed a generally decreasing trend. Our finding of a generally negative NMR for scholars is consistent with the results of studies by Schiller et al., which compared the birthplaces and workplaces of researchers, and reported that Germany has suffered a net loss of 28% of researchers due to international mobility (Schiller & Revilla Diez, 2008; Schiller & Cordes, 2016). The recent trends in migration among scholars contrasts with the NMR in the general population, which indicates that since 2008, Germany has increasingly become an immigrant destination. Indeed, since 2012, Germany has admitted large numbers of asylum-seekers.

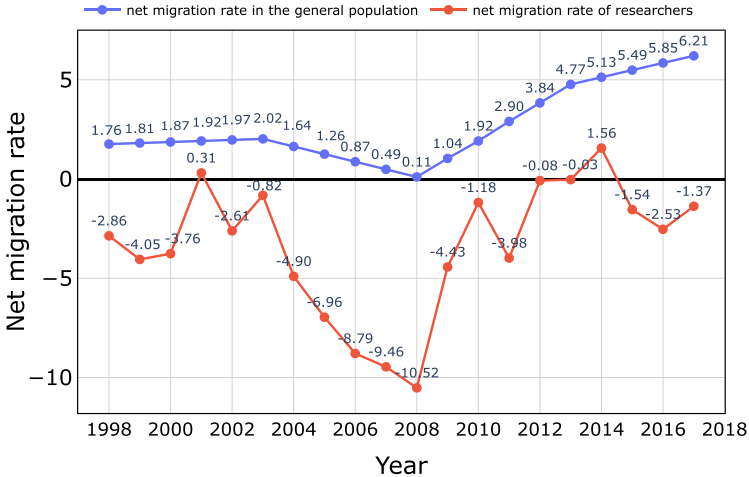

**Figure 3. Net migration rates for researchers (in red) and for the general population (in blue).**

*Citation-based performance by geography*

We previously examined the disparities in the citation-based performance of researchers of different mobility types (Fig. 1). In this subsection, we analyze the citation performance of immigrant and emigrant researchers by country. First, we divide the annual citation rate of each migrant researcher by the average of all migrant researchers in the respective discipline to obtain a discipline-normalized annual citation measure. Based on the three quantiles of the resulting distribution for all migrant researchers, we have distinguished three equally sized citation groups: low, medium, and high. Accordingly, the discipline-normalized annual citation rate is between 0.18 and 0.73 for migrant researchers in the medium group, it is below 0.18 for those in the low group, and it is above 0.73 for those in the high group.

Figure 4 shows the composition of the citation groups among the emigrants and immigrants for 30 countries.

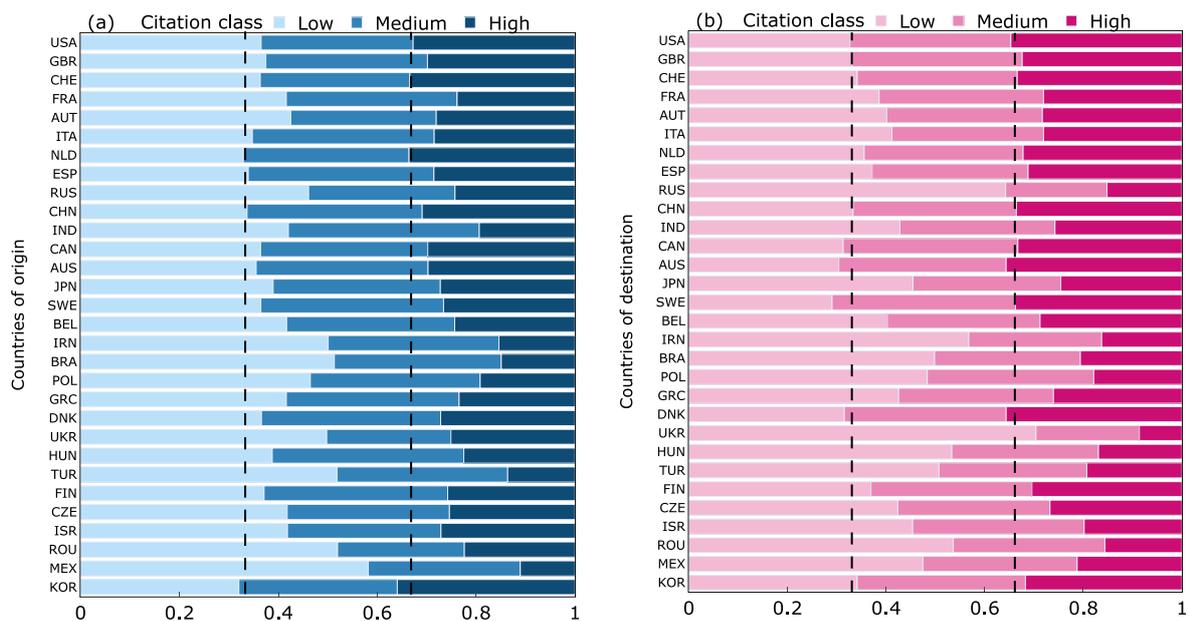

**Figure 4. Citation class composition of immigrants by origin (a) and emigrants by destination (b). Vertical gridlines are shown to enable a comparison to the average migrant researcher, who is equally likely to belong to any of the three citation classes.**

Among emigrants, scholars moving to Denmark, Austria, and Sweden are more likely than other emigrants to belong to the high citation group. Among the common destinations for emigrants, Sweden, Austria, and Denmark rank ninth, 10th, and 13th, respectively. Among immigrants, those who come from South Korea and the Netherlands are more likely than those who come from other countries to belong to the high citation group.

When we look at the numbers of immigrant and emigrants from the 10 countries with the largest flows, we observe that for most citation groups and country combinations, there are more emigrants than immigrants. This finding is also consistent with the overall negative net migration rate that we observed in Figure 3. For the US, the UK, and Switzerland (which have the three largest flows), the number of emigrants is considerably higher than the number of immigrants. Conversely, while Russia, Italy, and Spain, have comparable migration flows with Germany, they send more published researchers to Germany than they receive from Germany.

*Gender disparities among migrant researchers by discipline*

Figure 5 shows on the X-axis that almost all disciplines in Germany are dominated by male researchers. Especially for the disciplines in the physical sciences, like engineering, physics, and astronomy, the male-to-female ratio exceeds eight. The only disciplines for which the gender ratio is close to unity are veterinary sciences, psychology, and immunology and microbiology.

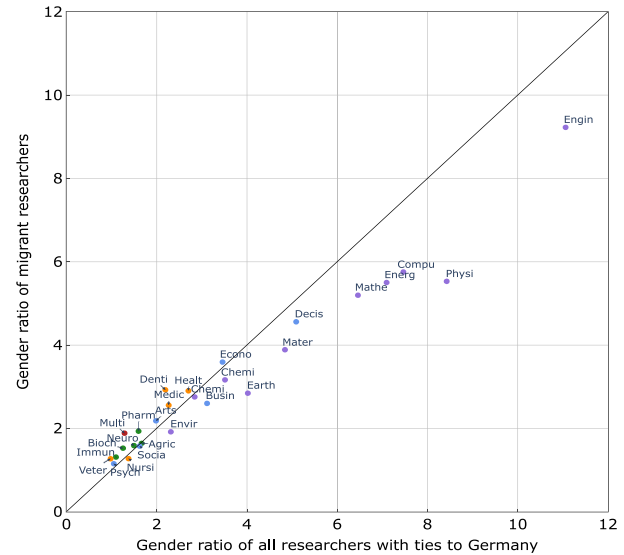

Given these observations, we examine the gender ratios among the migrant researchers of each discipline, and then compare them to the corresponding ratios among all German-affiliated researchers in Figure 5. The most striking finding is that in disciplines that have a gender ratio above four among all researchers, the gender disparity is less severe among the migrant researchers (see the data points under the 45° line, and especially all of the disciplines with ratios above four). In contrast, for most disciplines in the life sciences (green dots), the male-to-female ratio is more balanced than it is in the physical sciences, and it is slightly higher among migrant researchers.

**Figure 5. Male-to-female gender ratios of migrants and all researchers by discipline**

**Discussion and future directions**

As researchers play a unique role in innovation and economic growth, and are among the most mobile members of the occupational structure, most countries impose few, if any, limits on their mobility. Thus, it is crucial that we understand whether and, if so, how countries like Germany are situated in the global migration flows of published researchers (Docquier & Marfouk, 2004; Schiller & Cordes, 2016; Aref et al., 2019). We used Scopus bibliometric data (Falagas et al., 2008; Mongeon & Paul-Hus, 2016) to extrapolate the key variables and to infer the geographic locations of researchers. In addition, we drew from these data information on individual researchers, including on their scientific discipline, gender, and citation performance. Based on this information, we created five distinct mobility categories for researchers who had multiple publications and were exposed to mobility. These categories accounted for 55.99% of the 1.4 million author profiles identified in the Scopus data.

Algorithmically pre-processed and author-disambiguated Scopus data enabled us to provide uniquely rich, albeit preliminary, insights into Germany's role in the global system of migration among published researchers. Examining the 24-year period from 1996 to 2020, we demonstrated that internationally mobile researchers accounted for nearly 14% of the population of published researchers with ties to Germany. This group of internationally mobile researchers made disproportionately large contributions, as their differences in citation-based performance compared very favorably to those of non-movers in Germany.

The sheer size of our census of German-affiliated researcher data allowed us to differentiate the population of internationally mobile researchers according to their countries of origin and destination. Thus, these data provided us with a unique and detailed mapping of the key countries associated with academic mobility to and from Germany. We found that the US, the UK, and Switzerland were the three largest origin countries of published researchers immigrating to Germany. Our findings also showed that Germany has gone beyond reciprocating these levels of migration, with higher numbers of researchers emigrating from Germany to these countries. While these three countries were the primary destinations and accounted for 58% of all scholarly migration in this period, they were not necessarily the most significant in terms of their citation performance. Researchers who emigrated to Denmark, Sweden, and Austria and researchers who immigrated from South Korea and the Netherlands were more likely to be highly cited than other German-affiliated immigrants and emigrants.

Our data also allowed us to explore levels of gender disparities in the German science system, to investigate how these inequalities varied between migrant and *stationary researchers* (non-movers and single-paper authors), and to explore the heterogeneity in levels of gender inequality across scientific fields. We observed that the gender ratios (males to females) were nearly equal in disciplines such as veterinary sciences and psychology, whereas in fields such as engineering and physics and astronomy, there were six to eight times more men than women. While some of these patterns have been well-documented in the literature (Larivière et al., 2013; Macaluso et al., 2016; Sugimoto et al., 2015; IDEA Consult et al., 2017), our results highlight the key role that international mobility may play in helping to moderate some of the most extreme gender disparities. We found that female researchers were better represented among migrants than among stationary researchers in the most gender-imbalanced disciplines. This observation allows us to speculate that migration could further counterbalance the extremely gendered nature of certain disciplines in Germany. The results also highlight that more can be done to support female scholars wishing to remain in Germany, and to encourage the immigration of female scholars in fields that are heavily unequal, such as engineering, physics and astronomy, computer science, energy, and mathematics.

The analysis enabled us to uncover – within the quality limitations of the bibliometric data and our pre-processing approach for handling missing values and disambiguating authors – new aspects of the academic life course, and to show how it is connected to scholarly migration. Ultimately, our analysis helps to shed light on Germany as both a hub and a through station on the international map of academic mobility. The preliminary questions explored here offer some initial insights for policymakers, as they improve our understanding of the role of migrant researchers in the German science system, and their demographic composition. Our analysis contributes to the science of science by providing relevant evaluation measurements of the mobility of researchers from all disciplines based on their geography, gender, and citation performance.

Our study opens a number of new directions for further research. Building on this work, we intend to move from using an aggregate perspective to applying a micro-level perspective in order to gain further insights into the relationship between migration and scientific performance. This approach will be particularly useful for studying how women are affected, especially in fields where they are severely under-represented, and for examining the impact of specific national policies in Germany that aim to improve the gender balance in academic fields. Ultimately, by understanding the role of migration both into and out of Germany in each field and discipline, we can provide additional insights into researchers' characteristics, mobility behavior, and performance. Thus, our study represents a key step toward gaining a better understanding of migration among scholars.

## Acknowledgments

The authors highly appreciate the technical support from Tom Theile and the comments from Aliakbar Akbaritabar. This study has been funded by the German Academic Exchange Service with funds from the Federal Ministry of Education and Research. This study has received access to the bibliometric data through the project "Kompetenzzentrum Bibliometrie," and the authors acknowledge their funder BMBF (funding identification number 01PQ17001).